\documentclass[aps,prd,preprint,groupedaddress]{revtex4}

\usepackage{epsfig}
\usepackage{graphicx}
\usepackage{dcolumn}
\usepackage{bm}
\usepackage{amsmath}
\usepackage{slashbox}

\begin{document}

\def\beq{\begin{equation}}
\def\eeq{\end{equation}}
\def\bea{\begin{eqnarray}}
\def\eea{\end{eqnarray}}

\title{Fragmentation Functions for $K^0_S$ and $\Lambda$ with Complete
Quark Flavour Separation}
\author{S. Albino}
\affiliation{Institut f\"ur Theoretische Physik und Astrophysik,\\ 
Universit\"at W\"urzburg, 97074 W\"urzburg, Germany}
\author{B. A. Kniehl}
\affiliation{{II.} Institut f\"ur Theoretische Physik, Universit\"at Hamburg,\\
             Luruper Chaussee 149, 22761 Hamburg, Germany}
\author{G. Kramer}
\affiliation{{II.} Institut f\"ur Theoretische Physik, Universit\"at Hamburg,\\
             Luruper Chaussee 149, 22761 Hamburg, Germany}
\date{\today}
\begin{abstract}
We present new sets of next-to-leading order fragmentation functions
for the production of $K^0_S$ and $\Lambda$ particles from the gluon
and from each of the quarks, obtained by fitting to all relevant
data sets from $e^+ e^-$ annihilation. The  
individual light quark flavour fragmentation
functions are constrained phenomenologically for the first time by
including in the data the light quark tagging probabilities
measured by the OPAL Collaboration.
\end{abstract}


\maketitle


\section{Introduction}
\label{Intro}

At present, experimental results on inclusive hadron production from $e^+ e^-$
collisions are
the most reliable source for the extraction of universal fragmentation functions (FFs) $D_a^h(x,Q^2)$
(where $a$ labels the fragmenting parton, $h$ labels the produced hadron, $x$ is the fraction
of the parton's momentum taken by the produced hadron and $Q$ is the factorization scale),
which are crucial for making predictions for such processes in future
experiments, as well as for understanding the non-perturbative mechanism of hadron formation
in parton jets. However, the extraction of quark flavour
separated FFs from experimental data has not been completely possible due to the lack of data
for processes in which the individual light quark flavours are tagged. 
Therefore, theoretical assumptions between light
quark flavour FFs had to be made. Recently, we published sets of FFs \cite{Albino:2005me}
for each of the three light charged hadrons,
whose quark flavours were completely phenomenologically separated by including 
the light quark flavour separated measurements from the
OPAL collaboration \cite{Abbiendi:1999ry} in the data used for the fitting. This more reliable
separation in the light quark flavour sector of the FFs via real experimental
data is important for the description of
hadron production in proton-(anti)proton collisions, for example at the
RHIC, Tevatron, LHC and other experiments, because the proton is composed
predominantly of light partons.

In this paper, we extend our analysis of Ref.\ \cite{Albino:2005me} to determine
FFs for $K^0_S$ and $\Lambda$ production, in which the quark flavours are phenomenologically separated
by including,
among the available data, the tagging probability measurements for each of the quark 
flavours for these two particles provided
by OPAL in Ref.\ \cite{Abbiendi:1999ry}.
FFs for $K^0_S$ production have been previously obtained in Ref.\ 
\cite{Binnewies:1995kg}, however since no data was available to separate
the light quark flavours,
it was assumed that the $d$ and $s$ quark FFs were equal.
Much data for $K^0_S$ production can also be well described by using
FFs for $K^{\pm}$ \cite{Kniehl:2000fe,Albino:2005me}, so it is interesting
to verify if these two sets of FFs are really consistent.
In Ref.\ \cite{deFlorian:1997zj}, FFs for $\Lambda$ production were obtained by
constraining all light quark flavours to be equal and imposing certain
relations between them and the heavy quark FFs, since no data was available at the time to
fully constrain the individual quark flavour FFs. 
A further determination of FFs for $\Lambda$ production was obtained more recently
in Ref.\ \cite{Bourrely:2003wi}, along with FFs for the other 
octet baryons, again with relations imposed between flavours, but also
with some assumptions suggested by fermionic and bosonic statistics. 
A more reliable determination of these 
FFs is important due to the recent data on $K^0_S$ and $\Lambda$ production, taken
by the STAR Collaboration
\cite{Heinz:2005pg} at BNL RHIC. In particular, next-to-leading order (NLO) 
calculations deviate considerably
from the data for $\Lambda$ production.
For the first time, we present FFs for these particles without imposing
constraints on the non-perturbative components (except for the choice of parameterization
for the FFs).

\section{Method}

In all cross section calculations in this paper, used to 
fit FF parameters to data and to produce comparisons to this data and other
data, we use precisely the same method and choice of parameterization, scales etc.\
as in Ref.\ \cite{Albino:2005me}, 
and therefore we refer the reader
to this paper for details. This includes taking the fitted NLO value 
$\Lambda_{\overline{\rm MS}}^{(5)}=221$ MeV
of Ref.\ \cite{Albino:2005me}. For each parton, 
our FFs for $K^0_S$ production are defined to be those for the production of
a single $K^0_S$ particle, which is equal to the average of those for 
$K^0$ and $\overline{K}^0$, and 
our FFs for $\Lambda$ production are defined
to be the sum of those for $\Lambda^0$ and $\overline{\Lambda}^0$. 
Our FFs incorporate both the intrinsic (non-perturbative) and 
extrinsic (perturbative, dynamically generated) components of hadron production.
It is important to note that our FFs
contain intrinsic transitions involving intermediate hadrons 
occuring over durations much greater than that of the interaction, for example the process
$q+\overline{q}\rightarrow \Sigma^0 +X \rightarrow \Lambda +\gamma +X$ in $\Lambda$ production
noted in Ref.\ \cite{deFlorian:1997zj}, since such processes are not subtracted from the data.
Treatment of such effects are beyond the scope of this work.

To obtain FFs for $K_S^0$ production, we fit to all available $e^+ + e^- \rightarrow K_S^0 +X$ 
data, covering a range of centre-of-mass energies $\sqrt{s}$,
being the untagged data from TASSO at $\sqrt{s}=$14, 22 and 34 GeV \cite{Althoff:1984iz} 
and at 14.8, 21.5, 34.5, 35 and 42.6 GeV \cite{Braunschweig:1989wg},
from HRS \cite{Derrick:1985wd}, MARK II \cite{Schellman:1984yz} and TPC \cite{Aihara:1984mk} at 29 GeV,
from TASSO at 33.3 GeV \cite{Brandelik:1981ta}, from CELLO at 35 GeV \cite{Behrend:1989ae},
from TOPAZ at 58 GeV \cite{Itoh:1994kb}
and from ALEPH \cite{Barate:1996fi}, DELPHI \cite{Abreu:1994rg}, OPAL \cite{Abbiendi:2000cv} 
and SLD \cite{Abe:1998zs} at 91.2 GeV. The data from DELPHI at 183 and 189 GeV \cite{Abreu:2000gw}
are not included in the fit as will be explained below.
In addition, light, $c$ and $b$ quark tagged data from SLD at 91.2 GeV \cite{Abe:1998zs}
and $u$, $d$, $s$, $c$ and $b$ quark tagged probability measurements from OPAL at 91.2 GeV
\cite{Abbiendi:1999ry} are used to separate the individual quark flavours of the FFs.
Likewise, FFs for $\Lambda$ production are obtained by fitting to all 
available $e^+ + e^- \rightarrow \Lambda^0(\overline{\Lambda}^0) +X$ 
untagged data from TASSO at 14, 22 and 34 GeV 
\cite{Althoff:1984iz}, 
from HRS \cite{Baringer:1986jd} and MARK II \cite{delaVaissiere:1984xg} at 29 GeV,
from TASSO at 33.3 GeV \cite{Brandelik:1981ta} and at 
34.8 and 42.1 GeV \cite{Braunschweig:1988wh},
from CELLO at 35 GeV \cite{Behrend:1989ae},
from ALEPH \cite{Barate:1996fi}, DELPHI \cite{Abreu:1993mm}, OPAL \cite{Alexander:1996qj}
and SLD \cite{Abe:1998zs} at 91.2 GeV and from DELPHI at 183 and 189 GeV \cite{Abreu:2000gw}.
The individual quark flavours of the FFs are separated using
light, $c$ and $b$ quark tagged data from SLD at 91.2 GeV \cite{Abe:1998zs}
and $u$, $d$, $s$, $c$ and $b$ quark tagged probability measurements from OPAL at 91.2 GeV \cite{Abe:1998zs}.
Soft gluon effects \cite{Dokshitzer:1991wu} 
cause the DGLAP evolution to fail at small $x=2p/\sqrt{s}$, where $p$ is the
momentum of the observed hadron produced in the final state. Consequently we restrict our analysis
to data for which $x>0.1$, which implies that our FFs will not be valid below 
this $x$ value. Thus we have a total of 193 data points for $K_S^0$ production and
of 129 for $\Lambda$ production.
The individual systematic errors on the data sets are not given in the literature
by these experimental collaborations, which means that, like all previous FF determinations,
we are limited to fixing the off-diagonal elements of the covariance matrices for the data sets
to zero. We stress however that the experimental constraints on FFs (and $\alpha_s(M_Z)$)
would be greatly improved by the inclusion of these correlation effects.

\section{Results}

In this section
we perform fits to $K_S^0$ and $\Lambda$ production data to determine FFs for these 
particles.
In a first analysis, the FF parameters $N$, $\alpha$ and $\beta$ for the gluon and each of the
5 quark flavours are released. While an overall good fit is obtained, the fitted
initial gluon FFs are negative for both particles. Such a result may be unphysical.
Redoing the fits with the initial gluon FFs fixed to zero (so that the evolved
gluon FF in each case is generated dynamically from the quarks) gives similar results,
implying that the initial gluon FFs are consistent with zero with respect to these data. 
This is to be expected since the
gluon FF only enters the cross section at NLO, and can therefore only be reasonably
constrained by more accurate data such as that in
Ref.\ \cite{Albino:2005me}. To obtain a set of FFs in which the gluon FF is reliable,
we redo both fits with the initial gluon FF for each particle fixed to
some function whose choice and motivation
will now be explained. Using the approximation
\beq
D_a^{K_S^0}=\frac{1}{2}D_b^{K^{\pm}},
\label{ffforks0tokpm}
\eeq
where $b=u,d$ if $a=d,u$, otherwise $b=a$, we fix the inital gluon FF for $K_S^0$ production 
to half the AKK \cite{Albino:2005me} initial gluon FF for $K^+ + K^-$ production.
For $\Lambda$ production, it is reasonable to assume that the gluon FF
is related to that of the proton. However, since the valence structure of 
$\Lambda$ is $uds$ compared to $uud$ for the proton, we assume that this relation
holds with an overall suppression factor, which we determine approximately by
requiring a good description of the
$pp$ initiated cross section measurements from the STAR collaboration \cite{Heinz:2005pg}
to be discussed later.
We find that a good description of such data is achieved if the initial gluon FF
for $\Lambda$ production is fixed in the fit to the AKK initial gluon FF for $p +\overline{p}$
production suppressed by a factor of 3. Thus the STAR data 
is in some sense crudely included in the fit, at least in the sense of providing 
some constraint on the gluon FF for $\Lambda$ production.

We obtain $\chi^2_{\rm DF}=1.14$ and $\chi^2_{\rm DF}=1.39$ for the fits to the 
$K^0_S$ and $\Lambda$ production
data, respectively, indicating that both fits are good overall. The quality of the fit to each
individual data set is determined by its $\chi^2_{\rm DF}$ value, and these are listed in Tables
\ref{chiDFforetas1} -- \ref{allchisL}.
The values in Tables \ref{chiDFforetas1} and \ref{chiDFforetas2} show that the agreement 
with the OPAL tagging probabilities is good.
The good agreement with heavy quark tagged data
is surprising when compared with our findings of Ref.\ \cite{Albino:2005me}.
For that analysis of light charged hadron production data, 
although we found good agreement with the DELPHI, SLD and TPC heavy quark tagged data,
a poor description of the OPAL heavy quark tagged data 
was obtained, which, as we suggested,
may be due to large angle gluon emission effects. Whatever effect caused this disagreement
is clearly not as significant for the $K_S^0$ and $\Lambda$ production data
we are considering here, although we note that we find a rather large 
$\chi^2_{\rm DF}$ value for $b$ quark tagged $K_S^0$ production.
The $\chi^2_{\rm DF}$ values
for $K_S^0$ production in Table \ref{allchisK} indicate that each data set is well fitted except
the DELPHI data at 183 and 189 GeV which are excluded from the fit. When these data are included,
their $\chi^2_{\rm DF}$ values remain high (and not much less than those shown in
Table \ref{allchisK}). Such data require further consideration.
In any case, only 3 data points of these DELPHI data lie in the region $x>0.1$, so
they do not significantly change the size of the data sample used in our fit.
We will therefore exclude them in our fit. Table \ref{allchisL} shows that 
unsatisfactory agreement was obtained only with untagged data from DELPHI and $b$ quark tagged
data from SLD, both at $\sqrt{s}=91.2$ GeV. The data 
points from DELPHI at 183 GeV (1 point) and 189 GeV 
(1 point) for $\Lambda$ production were excluded for the same reasons as in the
case of $K_S^0$ production, 
although the point at 183 GeV could be well fitted.

The results for the FF parameters are shown in Table \ref{pars}.
The value for $\alpha$ and/or $\beta$ may become large because
the shape of the FF at small and/or large $x$ respectively is not well constrained.
However, the value of $N$ is then large to compensate for this.
For $\Lambda$ production, the inital FF for the $d$ quark is negative,
probably because this initial FF is consistent with zero with respect to the data which constrains it.
The initial $u$ quark FF has a large negative $\alpha$, due to a lack of data to constrain
this FF at small $x$.

\begin{figure*}[ht!]
\centering
\setlength{\epsfxsize}{13.5cm}
\begin{minipage}[ht]{\epsfxsize}
\centerline{\mbox{\epsffile{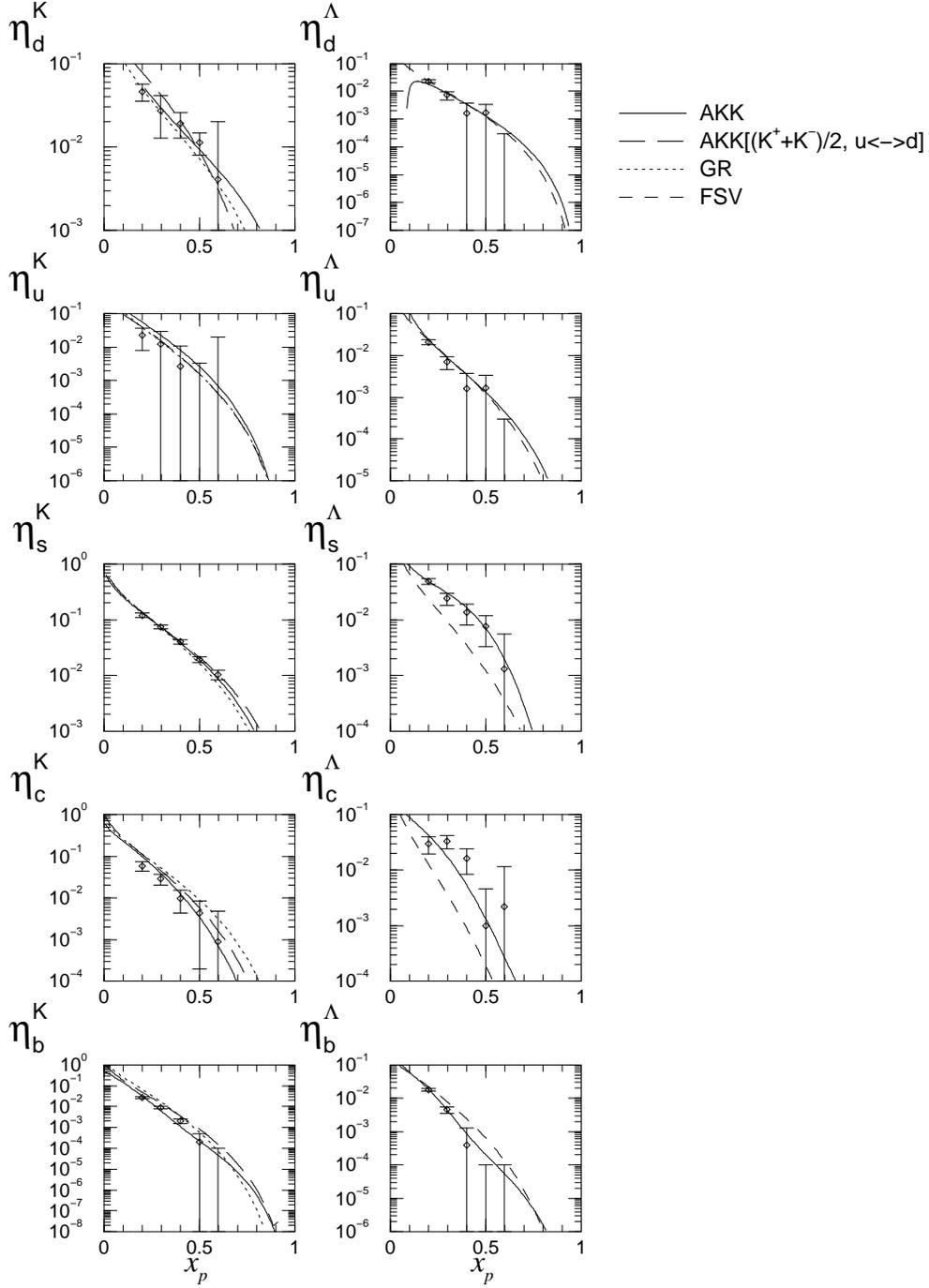}}}
\end{minipage}
\caption{
Quark tagging probabilities $\eta^h_a(x_p,s)$ at $\sqrt{s}=91.2$ GeV. 
The solid curves labelled ``AKK'' are calculated from our FF set for $K_S^0$ production
in the graphs on the left hand side and our FF set for $\Lambda$ production
in the graphs on the right hand side. The 
curves of the same label from the $K^{\pm}$ FF set in Ref.\ \cite{Albino:2005me} (after interchanging
$u$ and $d$) are also shown. In addition, we show the prediction for $K_S^0$ production from the 
$K_S^0$ FF set of
Ref.\ \cite{Greco:1994bn} (labelled ``GR'') and the prediction for $\Lambda$ production
from the $\Lambda$ FF set of Ref.\ \cite{deFlorian:1997zj} (labelled ``FSV''). 
The corresponding measured OPAL probabilites of
Ref.\ \cite{Abbiendi:1999ry} are also shown.
\label{fig1}}
\end{figure*}

\begin{figure*}[ht!]
\centering
\setlength{\epsfxsize}{11.5cm}
\begin{minipage}[ht]{\epsfxsize}
\centerline{\mbox{\epsffile{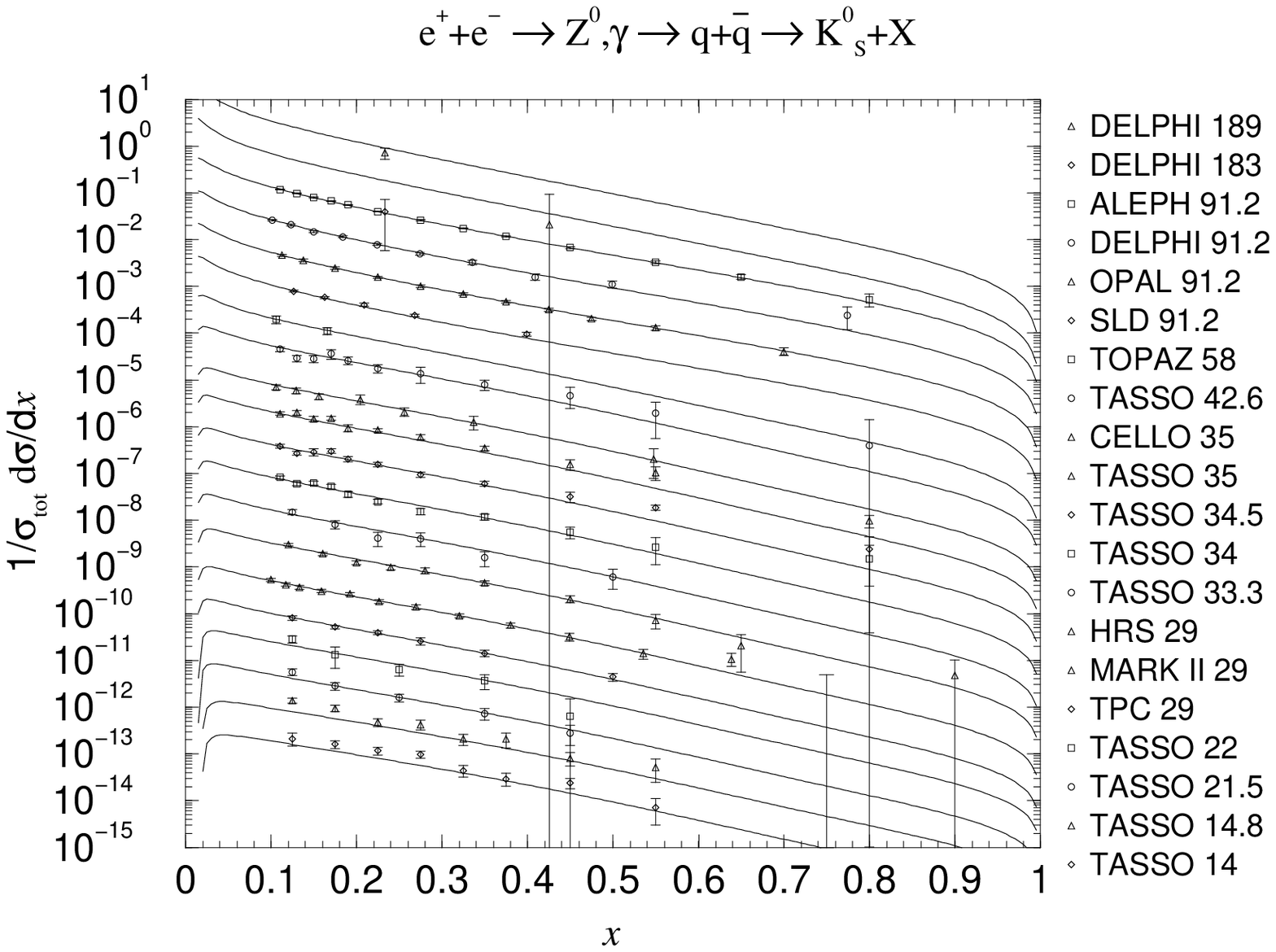}}}
\end{minipage}
\caption{
Normalized differential cross section of inclusive $K^0_S$ production.
The curves are calculated from the FFs obtained in our analysis, at the various energies
of the data shown. Each curve and the corresponding data
are rescaled relative to the nearest upper one by a factor of 1/5.
\label{fig2}}
\end{figure*}

\begin{figure*}[hb!]
\centering
\setlength{\epsfxsize}{11.5cm}
\begin{minipage}[ht]{\epsfxsize}
\centerline{\mbox{\epsffile{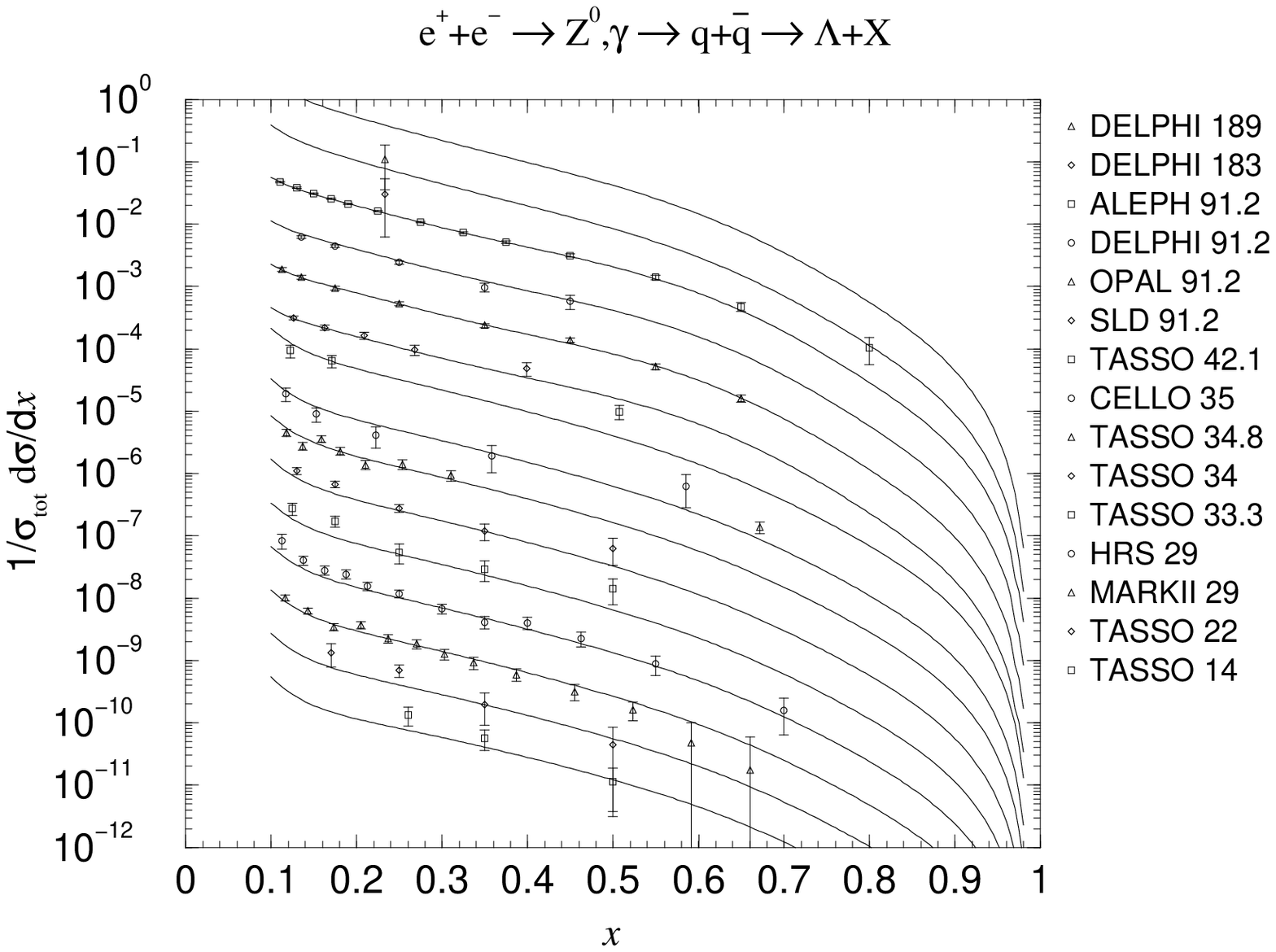}}}
\end{minipage}
\caption{
Normalized differential cross section of inclusive $\Lambda$ production.
The curves are calculated from the FFs obtained in our analysis, at the various energies
of the data shown. Each curve and the corresponding data
are rescaled relative to the nearest upper one by a factor of 1/5.
\label{fig3}}
\end{figure*}

The graphical comparison with the quark tagging probabilites is shown in 
Fig.\ \ref{fig1}. The curves obtained from our FF set in this analysis (labelled ``AKK'')
agree well with the data. 
For $K_S^0$ production, we also show the curves obtained from the FF set for $K^{\pm}$
production in Ref.\ \cite{Albino:2005me}, assuming the relation in Eq.\ (\ref{ffforks0tokpm}).
These curves are consistent with these measurements.
Also shown are the curves obtained from the FF set for $K_S^0$ production
of Ref.\ \cite{Greco:1994bn} (labelled ``GR''). These curves are consistent with the 
data, showing the validity of these authors' method of 
obtaining these FFs from the Monte Carlo generator HERWIG. 
We note that curves calculated from the FFs for $K^{\pm}$ production from Ref.\ \cite{Greco:1994bn}
(also obtained from HERWIG)
using the same assumptions (Eq.\ (\ref{ffforks0tokpm}))
are also consistent with the $K_S^0$ quark tagging probabilites.
For $\Lambda$ production, we also show the
curves obtained from the FF set presented in Ref.\ \cite{deFlorian:1997zj} (labelled ``FSV'').
These are generally consistent with the data except for the $s$ and $c$ quark tagged data. 
Unfortunately, we are unable to show the predictions from the FF set for $\Lambda$ production
of Ref.\ \cite{Bourrely:2003wi}, since the Mellin transform of the parameterization
used therein, that is required for our Mellin inversion calculation of
the cross section \cite{Albino:2005me}, cannot be obtained in closed form. 
Such predictions require a NLO $x$ space program.
The comparison with the differential cross sections is shown 
in Figs.\ \ref{fig2} and \ref{fig3}. For $K_S^0$ production, the disagreement
of the calculation
with the DELPHI data point at 183 GeV can be seen, and, of the two points at 189 GeV, with the one 
that is the higher in $x$.
For $\Lambda$ production, the failure of the description of the data point at 189 GeV can be seen.
Note that the DELPHI data for $\Lambda$ production
at 91.2 GeV that gave a high $\chi^2_{\rm DF}$ has rather small errors. Finally, in Fig.\ \ref{fig3}
the cross section for $\Lambda$ production in the region $x<0.1$ is not shown since large negative
values were obtained there. This results from the large negative value
for $\alpha$ in the FF for the $u$ quark discussed earlier.

\begin{figure*}[ht!]
\centering
\setlength{\epsfxsize}{9cm}
\begin{minipage}[ht]{\epsfxsize}
\centerline{\mbox{\epsffile{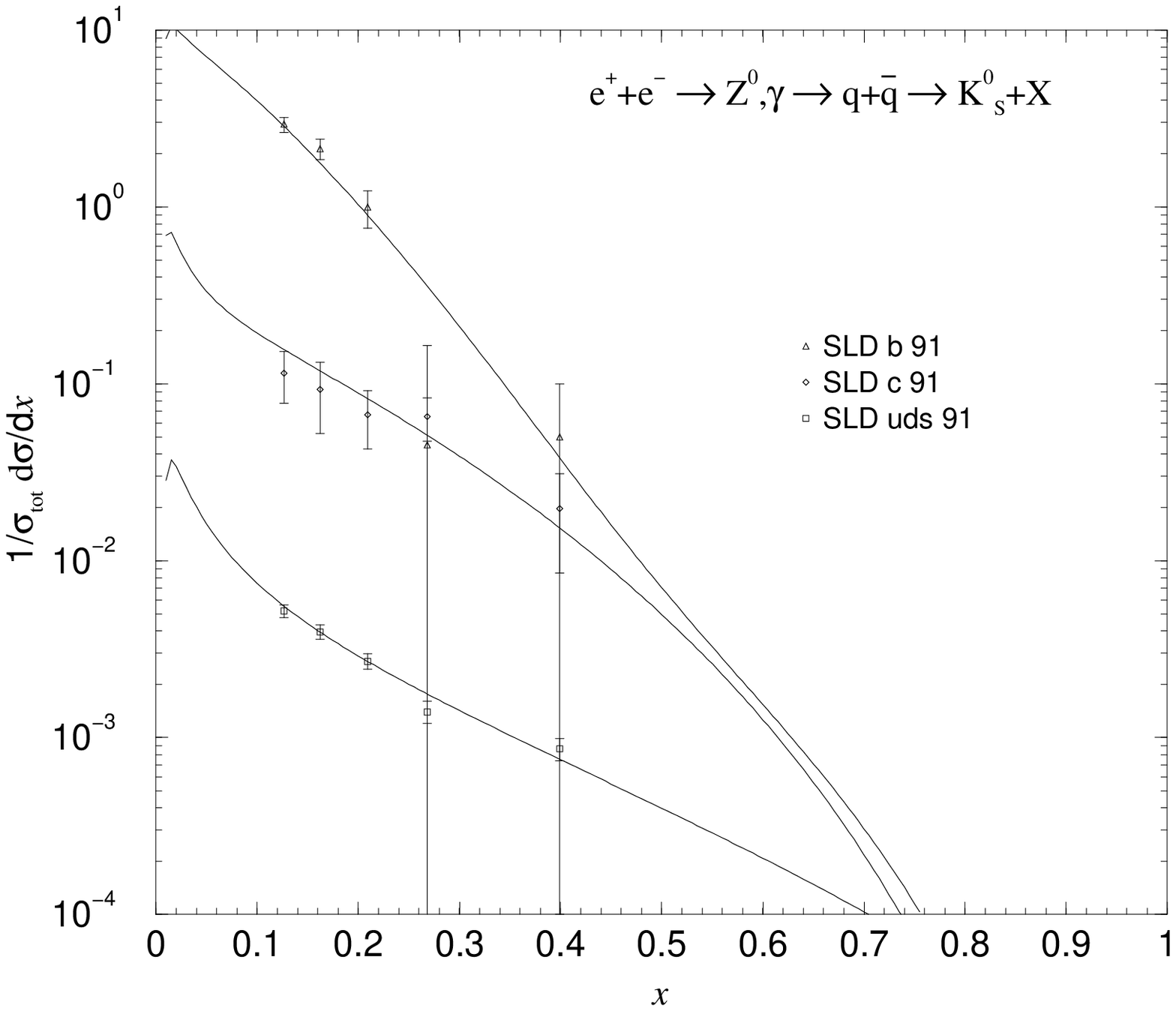}}}
\end{minipage}
\caption{
Normalized quark tagged differential cross sections for inclusive $K^0_S$ production
compared with corresponding data from SLD \cite{Abe:1998zs}.
The curves are calculated from the FFs obtained in our analysis. 
Each curve and the corresponding data
are rescaled relative to the nearest upper one by a factor of 1/20.
\label{fig4}}
\end{figure*}

\begin{figure*}[hb!]
\centering
\setlength{\epsfxsize}{9cm}
\begin{minipage}[ht]{\epsfxsize}
\centerline{\mbox{\epsffile{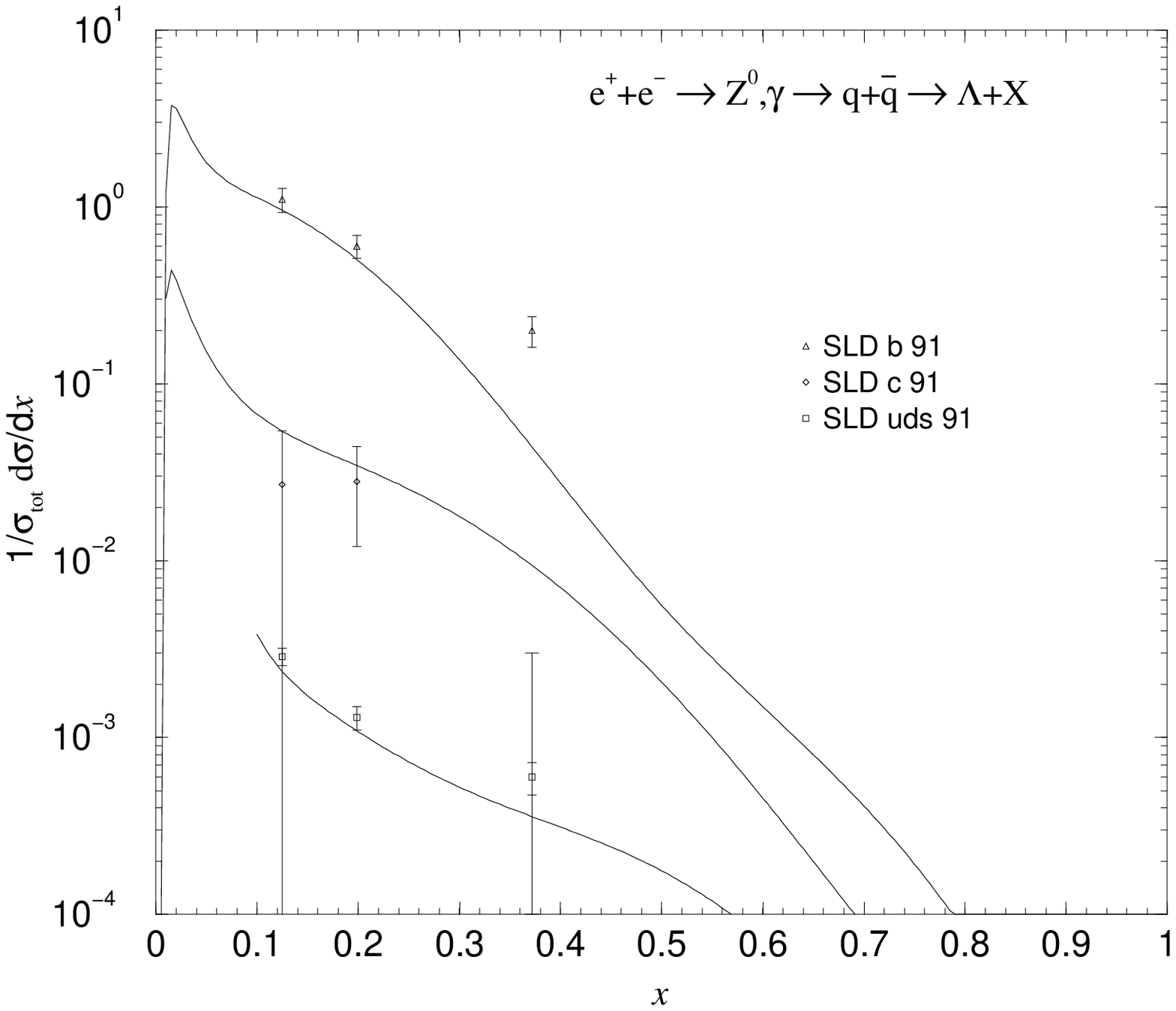}}}
\end{minipage}
\caption{
Normalized tagged differential cross sections for inclusive $\Lambda$ production
compared with corresponding data from SLD \cite{Abe:1998zs}.
The curves are calculated from the FFs obtained in our analysis. 
Each curve and the corresponding data
are rescaled relative to the nearest upper one by a factor of 1/20.
\label{fig5}}
\end{figure*}

The quark tagged data from SLD is shown in Figs.\ \ref{fig4} and \ref{fig5},
together with the calculation of these cross sections from the FF sets of this analysis.
Reasonable agreement is found with most of the data. However, in Fig.\ \ref{fig5}
there is significant disagreement with the interval from the $b$ quark tagged
cross section for $\Lambda$ production whose central
point is just below $x=0.4$. The light flavour quark tagged cross section
in this figure is large and negative in the region $x<0.1$, again due to the
large negative value for $\alpha$ in the FF for the $u$ quark.

We find that the large initial strange and charm quark FFs for $\Lambda$ production render the
other quark FFs negligible at large $x$, and this is reflected at higher
scale in the OPAL tagging probabilities at large 
$x_p$. This finding was not observed in Ref.\ \cite{deFlorian:1997zj}, where the light
quark FFs were constrained to be equal. This should make a significant difference
to hadron production data from proton-(anti)proton collisions, where the light quark
PDFs are important.

\begin{figure*}[ht!]
\centering
\setlength{\epsfxsize}{9cm}
\begin{minipage}[ht]{\epsfxsize}
\centerline{\mbox{\epsffile{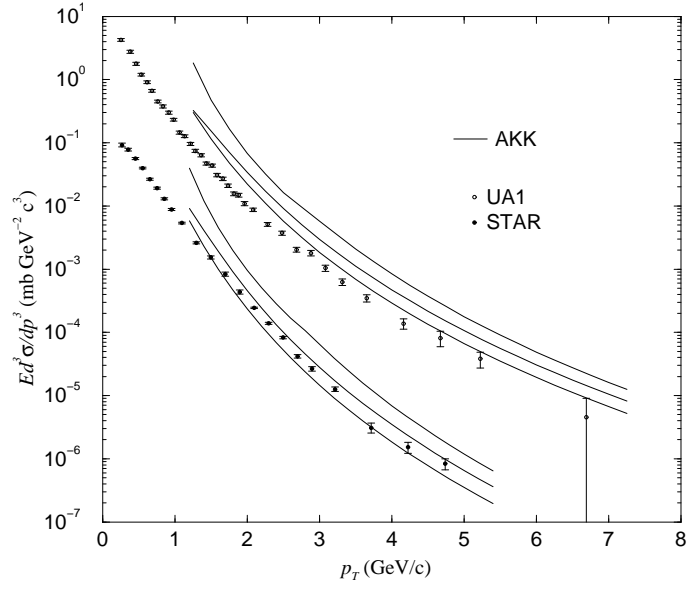}}}
\end{minipage}
\caption{
The invariant differential cross section for inclusive $K_S^0$ production
in $pp$ and $p\overline{p}$ collisions. Preliminary data from the STAR Collaboration
\cite{Heinz:2005pg} at $\sqrt{s}=200$ GeV (divided by a factor of 30 for clarity)
and data from the UA1 Collaboration \cite{Bocquet:1995jq} at $\sqrt{s}=630$ GeV are shown,
together with their predictions using the FFs obtained in
this paper. In the latter case, the upper, central and lower curves are calculated with
a renormalization and factorization scale of $\mu=M_f=p_T/2$, 
$p_T$ and $2p_T$ respectively.
\label{fig6}}
\end{figure*}

\begin{figure*}[hb!]
\centering
\setlength{\epsfxsize}{9cm}
\begin{minipage}[ht]{\epsfxsize}
\centerline{\mbox{\epsffile{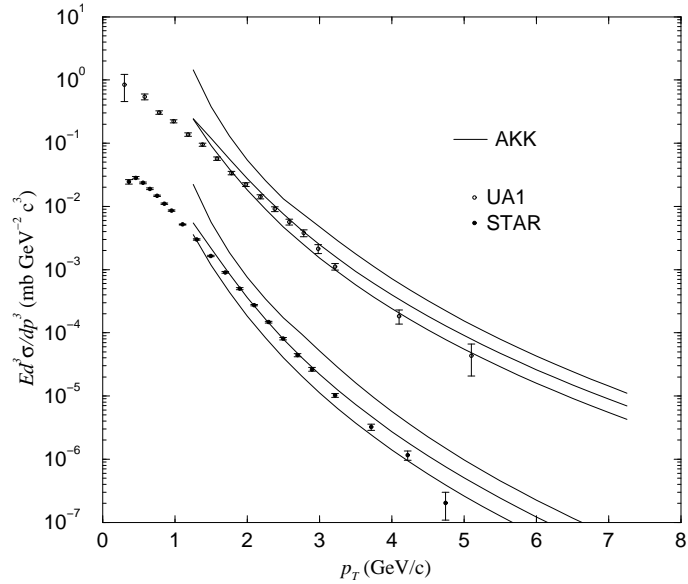}}}
\end{minipage}
\caption{
As in Fig.\ \ref{fig6}, but for $\Lambda$ production.
\label{fig7}}
\end{figure*}

To test our FFs, we use them to predict
the recent preliminary measurements for $pp$ initiated cross sections at $\sqrt{s}=200$ GeV 
from the STAR Collaboration \cite{Heinz:2005pg}
and for $p\overline{p}$ initiated cross sections at $\sqrt{s}=630$ GeV 
for $K^0_S$ and $\Lambda$ production from the UA1 Collaboration \cite{Bocquet:1995jq} at the CERN SPS.
We use the NLO coefficient functions for this process from Ref.\ \cite{ACGG}, and
the CTEQ6M parton distribution functions from Ref.\ \cite{Stump:2003yu}.
Figure \ref{fig6} shows the results for $K_S^0$ production, which
are very similar to those obtained in a similar plot in Ref.\ \cite{Albino:2005me}, as is to be expected
from the similarity between the two AKK curves of Fig.\ \ref{fig1}.
While good agreement with STAR data is found, some other effects are required for the 
description of the UA1 data. Figure \ref{fig7} shows the results for $\Lambda$ production.
As discussed earlier,
the STAR data were used to motivate the choice of the initial gluon FF: Firstly,
fixing the initial gluon FF to that of the AKK proton from Ref.\ \cite{Albino:2005me} and performing
a fit to the $e^+ e^-$ data, we found that the description of the STAR data was too high by a factor of 
about 3. Since we also found that these data are dominated by the gluon FF at low factorization scale,
we divided this choice of the initial gluon FF by this factor and redid the fit 
(i.e. fitted the initial quark FFs) to get 
our final FF set for $\Lambda$ production. The plot shows good agreement with both STAR and UA1 
data within the theoretical errors.

\section{Conclusions}

We have obtained FFs for $K_S^0$ and $\Lambda$ production by fitting
to data from $e^+ e^-$ collisions over a wide range of $\sqrt{s}$ values, from
14 to 91.2 GeV. To separate the light quark flavour FFs, and to improve the determination of the
heavy ones, we have included for the first time the quark tagging probabilities from Ref.\ 
\cite{Abbiendi:1999ry}. For $K_S^0$ production, we find good agreement
with the FFs from Ref.\ \cite{Greco:1994bn} and the FFs for 
$K^+ + K^-$ production from Ref.\ \cite{Albino:2005me} (using Eq.\ (\ref{ffforks0tokpm}) in the latter case)
adding support to the reliability of the data of Ref.\ \cite{Abbiendi:1999ry}.
For $\Lambda$ production, we found that the light quark flavour FFs are significantly different
at large $x$, with the initial strange quark FF dominating over the remaining quark FFs.

In both cases the initial gluon FF is fixed in the fit. For $K_S^0$ production, we use
the AKK initial gluon FF for $K^{\pm}$ (divided by 2). This leads to a good description
of the $pp$ initiated STAR data for $K_S^0$ production, but a bad description
of the UA1 data. This problem was also found in Ref.\ \cite{Albino:2005me} 
using the AKK FFs for $K^{\pm}$ production via Eq.\ (\ref{ffforks0tokpm}). 
For $\Lambda$ production, we use the AKK initial gluon FF
for $p/\overline{p}$ production, after dividing by a factor of 3. This factor is
chosen to give a good description of the STAR data for $\Lambda$ production, and also
leads to a good description of the UA1 data.
Since such data constrains the gluon well at low $p_T$, it should in future analyses
be directly included in the list of data fitted to. 
However, currently the calculation, which implements Monte Carlo integration,
is too slow to obtain fitted FFs in a reasonable time.

In order to make predictions, our fitted FF's over the range $0.1<x<1$ and $M_0<M_f<200$ GeV 
can be obtained from the FORTRAN routines at \verb!http://www.desy.de/~simon/AKK2005FF.html!, 
which are calculated using cubic spline interpolation from a linear grid in $x$
and linear interpolation from a linear grid in $\ln \sqrt{s}$.

\begin{acknowledgments}

The authors would like to thank M.\ Heinz for providing them with the preliminary
numerical values for the measurements of $K^0_S$ and $\Lambda$ production in $pp$
collisions from the STAR collaboration shown graphically in
Ref.\ \cite{Heinz:2005pg}. This work was supported in part by 
the Deutsche Forschungsgemeinschaft
through Grant No.\ KN~365/5-1 and by the Bundesministerium f\"ur Bildung und
Forschung through Grant No.\ 05~HT4GUA/4.

\end{acknowledgments}




\begin{table}[ht!]
\begin{footnotesize}
\renewcommand{\arraystretch}{1.1}
\caption{$\chi^2_{\mathrm{DF}}$ values
obtained from the measured quark tagging probabilities for $K_S^0$ production, $\eta_a^K$,
at $\sqrt{s}=91.2$ GeV in Ref.\ \cite{Abbiendi:1999ry}.\label{chiDFforetas1}}
\begin{center}
\begin{tabular}{||c||l|l|l|l|l||}
\hline\hline
$a$ & $d$ & $u$ & $s$ & $c$ & $b$ \\
\hline
$\chi^2_{\rm DF}$ &   0.27 &   1.38 &   0.54 &   1.77 &   2.36 \\
\hline\hline
\end{tabular}
\end{center}
\end{footnotesize}
\end{table}

\begin{table}[h!]
\begin{footnotesize}
\renewcommand{\arraystretch}{1.1}
\caption{$\chi^2_{\mathrm{DF}}$ values
obtained from the measured quark tagging probabilities for $\Lambda$ production, $\eta_a^{\Lambda}$,
at $\sqrt{s}=91.2$ GeV in Ref.\ \cite{Abbiendi:1999ry}.\label{chiDFforetas2}}
\begin{center}
\begin{tabular}{||c||l|l|l|l|l||}
\hline\hline
$a$ & $d$ & $u$ & $s$ & $c$ & $b$ \\
\hline
$\chi^2_{\rm DF}$ &   1.05 &   1.00 &   0.29 &   1.39 &   1.02 \\
\hline\hline
\end{tabular}
\end{center}
\end{footnotesize}
\end{table}

\begin{table*}[h!]
\begin{footnotesize}
\renewcommand{\arraystretch}{1.1}
\caption{CM energies, types of data and $\chi^2_{\mathrm{DF}}$ values
for various samples of $K_0^S$ production mesurements.
Samples not used in the fits are marked by asterisks. The columns are 
labelled by the quarks that were tagged in the mesurements ($q$ implies no tagging).
\label{allchisK}}
\begin{center}
\begin{tabular}{||c|l|l|l|l||}
\hline\hline
\backslashbox{$\sqrt{s}$ [GeV]}{tagged:} & $q$ & uds & $c$ & $b$ \\
\hline\hline
 14.0 &   1.82 \cite{Althoff:1984iz} & & & \\
 14.8 &   2.65 \cite{Braunschweig:1989wg} & & & \\
 21.5 &   0.69 \cite{Braunschweig:1989wg} & & & \\
 22.0 &   0.99 \cite{Althoff:1984iz} & & & \\
 29.0 &   1.27 \cite{Derrick:1985wd}   0.40 \cite{Schellman:1984yz}   0.16 \cite{Aihara:1984mk} & & & \\
 33.0 &   0.82 \cite{Brandelik:1981ta} & & & \\
 34.0 &   1.33 \cite{Althoff:1984iz} & & & \\
 34.5 &   1.89 \cite{Braunschweig:1989wg} & & & \\
 35.0 &   0.30 \cite{Braunschweig:1989wg}   1.64 \cite{Behrend:1989ae} & & & \\
 42.6 &   1.02 \cite{Braunschweig:1989wg} & & & \\
 58.0 &   0.01 \cite{Itoh:1994kb} & & & \\
 91.2 &   0.33 \cite{Barate:1996fi}   0.57 \cite{Abreu:1994rg}   0.31 \cite{Abbiendi:2000cv}   1.13  \cite{Abe:1998zs} &   0.93  \cite{Abe:1998zs} &   0.54  \cite{Abe:1998zs} &   1.85 \cite{Abe:1998zs} \\
183.0 &  21.87 \cite{Abreu:2000gw}$^*$ & & & \\
189.0 &   4.41 \cite{Abreu:2000gw}$^*$ & & & \\
\hline\hline
\end{tabular}
\end{center}
\end{footnotesize}
\end{table*}

\begin{table*}[hb!]
\begin{footnotesize}
\renewcommand{\arraystretch}{1.1}
\caption{CM energies, types of data and $\chi^2_{\mathrm{DF}}$ values
for various samples of $\Lambda$ production mesurements.
Samples not used in the fits are marked by asterisks. The columns are
labelled by the quarks that were tagged in the mesurements ($q$ implies no tagging).
\label{allchisL}}
\begin{center}
\begin{tabular}{||c|l|l|l|l||}
\hline\hline
\backslashbox{$\sqrt{s}$ [GeV]}{tagged:} & $q$ & uds & $c$ & $b$ \\
\hline\hline
 14.0 &   0.73 \cite{Althoff:1984iz} & & & \\
 22.0 &   0.73 \cite{Althoff:1984iz} & & & \\
 29.0 &   1.19 \cite{Baringer:1986jd}   0.83 \cite{delaVaissiere:1984xg} & & & \\
 33.0 &   1.59 \cite{Brandelik:1981ta} & & & \\
 34.0 &   1.85 \cite{Althoff:1984iz} & & & \\
 34.8 &   2.29 \cite{Braunschweig:1988wh} & & & \\
 35.0 &   0.88 \cite{Behrend:1989ae} & & & \\
 42.1 &   0.90 \cite{Braunschweig:1988wh} & & & \\
 91.2 &   0.43 \cite{Barate:1996fi}   4.05 \cite{Abreu:1993mm}   0.47 \cite{Alexander:1996qj}   0.33 \cite{Abe:1998zs} &   1.63 \cite{Abe:1998zs} &   1.54 \cite{Abe:1998zs} &   3.70 \cite{Abe:1998zs} \\
183.0 &   4.89 \cite{Abreu:2000gw}$^*$ & & & \\
189.0 &  16.34 \cite{Abreu:2000gw}$^*$ & & & \\
\hline\hline
\end{tabular}
\end{center}
\end{footnotesize}
\end{table*}

\begin{table*}[hb!]
\begin{small}
\renewcommand{\arraystretch}{1.1}
\caption{\label{pars}Values and errors of $N$, $\alpha$ and $\beta$ resulting from the fit.}
\begin{center}
\begin{tabular}{||c|l||c|c|c||}
\hline \hline 
Hadron & Flavour & $N$ & $\alpha$ & $\beta$ \\
\hline \hline 
$K^0_S$ 
& $d$ &     0.0297 &    -0.949 &     0.0953 \\
& $u$ &     1.29 &     0.137 &     4.53 \\
& $s$ &     0.560 &    -0.283 &     1.43 \\
& $c$ &     8.43 &     0.550 &     6.04 \\
& $b$ &    28.8 &     0.574 &    13.5 \\
& $g$ (fixed)&     7.96 &     2.72 &     2.45 \\
\hline \hline
$\Lambda$ 
& $d$ &   -16.7 &    -0.169 &    20.2 \\
& $u$ &     0.0016 &    -3.67 &     4.27 \\
& $s$ & 46043 &     9.27 &     9.11 \\
& $c$ &   156 &     2.71 &     9.51 \\
& $b$ &   244 &     2.15 &    15.6 \\
& $g$ (fixed)&     0.289 &     0.130 &     0.854 \\
\hline \hline
\end{tabular}
\end{center}
\end{small}
\end{table*}

\newpage

\newpage

\newpage

\newpage

\end{document}